Strain-induced enhancement of plasma dispersion effect and free-carrier absorption in SiGe optical modulators

Younghyun Kim[1*], Mitsuru Takenaka[1], Takenori Osada[2], Masahiko Hata[2], and Shinichi Takagi[1]

[1] *Department of Electrical Engineering and Information Systems, The University of Tokyo, 7-3-1 Hongo, Bunkyo-ku, Tokyo 113-8656, Japan*

[2] *Sumitomo Chemical Co. Ltd., 6 Kitahara, Tsukuba, Ibaraki 300-3294, Japan*

*E-mail: yhkim@mosfet.t.u-tokyo.ac.jp*

**Abstract**

The plasma dispersion effect and free-carrier absorption are widely used for changing refractive index and absorption coefficient in Si-based optical modulators. However, these free-carrier effects in Si are not large enough for making the footprint of the Si modulators small. Here, we have theoretically and experimentally investigated the enhancement of the plasma dispersion effect and free-carrier absorption by strain-induced mass modulation in silicon-germanium (SiGe). The application of compressive strain to SiGe reduces the conductivity hole mass, resulting in the enhanced free-carrier effects. Thus, the strained SiGe-based optical modulator exhibits more than twice modulation efficiency as large as that of the Si modulator. To the best of our knowledge, it is the first demonstration of the enhanced free-carrier effects in strained SiGe at the near-infrared telecommunication wavelength. The strain-induced enhancement technology for the free-carrier effects is expected to boost the modulation efficiency of the most Si-based optical modulators thanks to high complementary metal-oxide-semiconductor (CMOS) compatibility.

## Introduction

The plasma dispersion effect and free-carrier absorption are well known physical phenomena to change optical constants: refractive index (*n*) and absorption coefficient ($\alpha$) in semiconductors. In particular, these effects are the most promising for silicon (Si) to build-up optical modulators as previously shown in the study by Soref and Bennett[1]. Thus, Si optical modulators based on the free-carrier effects have been demonstrated thus far by carrier modulation through injection, depletion and accumulation[2-13]. Especially a depletion-based Mach-Zehnder interferometer (MZI) optical modulator, in which the refractive index of the phase shifters are modulated by carrier depletion, is one of the most promising modulators in terms of modulation speed and optical bandwidth. However, the weak plasma dispersion and free-carrier absorption in Si cause the low modulation efficiency, resulting in a device length of more than 1 mm for MZI modulators. To supplement the weak free-carrier effects in Si, the introduction of ring resonators instead of MZI[2,11] and slow-light structures using photonic crystal[12,13] have been investigated, while the optical bandwidth is also reduced in such structures. Hence, the enhancement of the plasma dispersion effect and free-carrier absorption will be fundamentally important to improve the performance of the Si-based optical modulators.

The plasma dispersion effects and free-carrier absorption are expressed by the Drude model, describing that changes in refractive index and absorption coefficient arise from change in a plasma frequency of free carriers which is dependent on not only the number of free carriers but also their conductivity effective masses. In the results, the refractive index change by the plasma dispersion effect is inversely proportional to conductivity effective masses of electrons and holes[14]. Therefore, the lighter conductivity masses become, the higher plasma dispersion occurs. In the CMOS technology, the application of strain to Si has been widely used for achieving lighter conductivity masses and higher mobilities in channels of transistors, which can overcome the fundamental difficulty in scaling the MOS transistors. Tensile strain and compressive strain is applied to Si n-channel MOS transistors and p-channel MOS transistors, respectively[15-17]. It was also experimentally observed that the plasma dispersion effect and free-carrier absorption in Si at the far-infrared wavelength range from 5 to 20 μm were modified by uniaxial strain mechanically applied to Si through strain-induced mass modulation [18-20]. Here, we have investigated the enhancement of plasma dispersion effect and free-carrier absorption by strain-induced mass modulation in Si/strained SiGe/Si double-heterostructure waveguide with a lateral *pin*-junction for carrier injection at the near-infrared wavelength range from 1.3 to 1.6 μm, which are most important wavelengths for optical communications. The Si/SiGe/Si device exhibits two times smaller injection carrier required for 20-dB attenuation than Si, which is well explained by the strain-induced enhancement of free-carrier absorption in SiGe. The enhancement of the plasma dispersion effect is also evaluated from the free-carrier absorption spectra through the Kramers-Kronig transform, which is well explained by the theoretical prediction according to the Drude model.

**Free-carrier effects in compressively strained SiGe**

Strain in SiGe coherently grown on Si is controllable by the Ge mole fraction of SiGe because the lattice constant of Ge is 4% larger than that of Si. Hence, the compressively biaxial strained SiGe film can be obtained when the thickness of SiGe grown on Si is less than its critical thickness[21,22]. The band structure of SiGe is highly affected by strain, resulting in mass modulations of electrons and holes. Figures 1 show the valance band structures of light hole (LH) and heave hole (HH) when the SiGe film with Ge fraction $x$ of 0, 0.15, and 0.3 are coherently grown on Si (001) as a function of the wave vector in the $k_x$ [100]- $k_y$ [010] plane calculated by $k{\cdot}p$ method[23-25]. In the case of Si ($x$ = 0), the HH band and LH band are degenerate at the Γ point. The compressive strain splits this degeneracy; thus, the HH band is shifted to above the LH band in the case of strained $Si_{1-x}Ge_x$ on Si ($x$ > 0). The energy surfaces of the LH and HH bands become sharp at the minimum energy Γ point, exhibiting that the biaxial compressive strain reduces the effective hole masses at the Γ point.

The conductivity effective masses of hole and electron as a function of Ge fraction are shown in Fig. 2. From the calculated valance band energies of LH and HH, the conductivity effective hole masses of strained and relaxed SiGe are obtained. The conductivity effective mass of electron is taken from the reference[26]. Comparing with the mass of Si ($x$ = 0), the conductivity hole mass of relaxed SiGe deceases with an increase in Ge fraction. Moreover, the conductivity hole mass dramatically decreases owing to the compressive strain. On the other hand, the in-plane mass of electron at Δ valleys stagnates regardless of Ge fraction in case of the application of biaxial compressive strain.

The plasma dispersion effect and free-carrier absorption are expressed by the Drude model, describing that changes in refractive index and absorption coefficient arise from change in a plasma frequency of free carriers which is dependent on the number of free carriers and their conductivity effective masses. Thus, the changes in refractive index ($\Delta n$) and absorption coefficient ($\Delta\alpha$) are expressed by

$$\Delta n = -(e^2\lambda^2/8\pi^2c^2\varepsilon_0 n)[\Delta N_e/m^*_{ce} + \Delta N_h/m^*_{ch}] \tag{1}$$

$$\Delta\alpha = (e^3\lambda^2/4\pi^2c^3\varepsilon_0 n)[\Delta N_e/m^*_{ce}{}^2\mu_e + \Delta N_h/m^*_{ce}{}^2\mu_h] \tag{2}$$

where $e$ is the electronic charge, $\varepsilon_0$ is the permittivity in vacuum, $c$ is the speed of light in vacuum, $\lambda$ is the wavelength, $n$ is the unperturbed refractive index, $m^*_{ce}$ and $m^*_{ch}$ are the conductivity effective masses for electron and hole, and $\mu_e$ and $\mu_h$ is the mobilities for electron and hole, respectively[1].

According to the Drude model, the changes in optical constants are inversely proportional to the conductivity effective mass for $\Delta n$ and the square of the conductivity effective mass multiplied by the mobility for $\Delta\alpha$. Therefore, the changes in the optical constants are expected to be enhanced by reducing conductivity effective masses.

Figures 3(a) and (b) show the enhancements of refractive index and absorption coefficient in compressively strained SiGe, in which the mobilities were taken from the literatures for electron[27] and for hole[28], respectively. The refractive index of SiGe was taken from the reference[29]. The changes in the optical constants by hole increase with an increase in Ge fraction because the conductivity hole mass of SiGe is reduced by compressive strain as shown in Fig. 2. However, there is no enhancement of the refractive index change by electron due to the almost no modulation in the electron mass by strain as shown in Fig. 2. The reduction in the electron mobility with an increase in Ge fraction slightly contributes the enhancement of the absorption coefficient change by electron, but much smaller than that by hole. When we consider a SiGe optical modulator into which carriers are injection through a forward-biased *pin*-junction, the same amount of electrons and holes exist in the SiGe layer due to the charge neutrality condition. Thus, the average of the changes in optical constants by injected electrons and holes in the *pin* structure are alleviated as shown in Figs. 3. However, the averaged enhancement factors of $\Delta n$ and $\Delta\alpha$ still increase with an increase in Ge fraction and are expected to be approximately 1.6 for $\Delta n$ and 2.1 for $\Delta\alpha$ in $Si_{0.7}Ge_{0.3}$.

**Numerical analysis of strained-SiGe based carrier-injection optical modulator**

We numerically analyzed the modulation characteristics of the carrier-injection waveguide optical modulator with the compressively strained SiGe well in the waveguide core by using the technology computer aided design (TCAD) simulation in conjunction with the finite-difference optical mode analysis. Figure 4 (a) presents the schematic of the device structure with a lateral *pin*-junction for carrier injection, which is fabricated on a silicon-on-insulator (SOI) substrate. The device structure of the strained SiGe-based optical modulator consists of a 115-nm-thick and 600-nm-width waveguide mesa with the constant doping level of $10^{16}$ $cm^{-3}$ for boron. The 30-nm-thick $Si_{1-x}Ge_x$ layer is embedded in the center of the waveguide mesa, which can be coherently grown on Si (001) when the Ge fraction is up to 0.3[22]. The $p^+$ and $n^+$ regions with the doping level of $10^{20}$ $cm^{-3}$ are formed at the both sides of the waveguide mesa for carrier injection. The fundamental transverse electric (TE) mode calculated by the finite-difference method is shown in Fig. 4(b). The optical confinement factor of the SiGe layer is estimated to be 19% by the finite-difference mode analysis. The carrier concentration is calculated by taking into account recombination processes such as Shockley-Read-Hall (SRH) and Auger recombination[30] when electrons and holes are injected by applying a forward-bias voltage between the $p^+$ to $n^+$ regions. We supposed ohmic contacts to the $p^+$ to $n^+$ regions. After calculating the carrier concentration, the changes in refractive index and absorption coefficient of SiGe and Si were calculated based on the Drude model described by Equations (1) and (2). Finally the changes in the effective refractive index and absorption coefficient were obtained by calculating overlap between the carrier distribution and the optical field distribution.

First, we show the carrier concentrations in Fig. 5(a) as a function of injected current density into the Si/SiGe/Si heterostructure modulator. It is found that the carrier concentration in the SiGe layer is much larger than that in the Si layer because the injected carriers are confined in the SiGe layer due to the band offset between Si and SiGe as shown in Fig. 5 (b), which also contributes to improve the modulation efficiency of Si-based optical modulators[4]. The carrier concentration in the SiGe layer increases by an order of magnitude as the Ge fraction increases from 0 to 0.3 because the band offset between Si and SiGe increases. Thus, it is possible to synergistically enhance the effective refractive index and absorption coefficient changes by gathering the injected carriers into the SiGe layer where the plasma dispersion effect and free-carrier absorption are larger than those of Si. Finally, the effective refractive index and absorption coefficient changes for the fundamental TE mode are shown in the Figs. 6 by considering synergetic two effects of the strain-induced mass modulation and the carrier confinement in SiGe layer. The solid lines in Fig. 6(a) show the optical attenuation characteristics with the Ge fraction from 0 to 0.3 as a function of injected current density. Owing to the free-carrier absorption enhancement and the carrier confinement in the SiGe layer, the optical attenuation increases with the Ge fraction. Figure 6(b) shows the refractive index change which has same trend as the attenuation characteristics. To identify the carrier confinement effect on the modulation efficiency, the modulation characteristics considering only the carrier confinement, in which the enhancements of the plasma dispersion effect and free-carrier absorption in SiGe are neglected, are calculated as shown in the broken lines in Figs. 6(a) and (b). Although the carrier confinement partly contributes to enhance the device performance, the strain effect plays the most important role for the enhancement of the modulation characteristics. Figure 6(c) shows the current density required for 20 dB attenuation and the enhancement factor of the optical attenuation against Si as a function of Ge fraction. Comparing with the Si case ($x$ = 0), the current density required for 20 dB-attenuation significantly decreases from 57 mA to 15 mA in the case of $Si_{0.7}Ge_{0.3}$. Hence, the efficiency of the $Si_{0.7}Ge_{0.3}$-based in-line intensity modulator based on the free-carrier absorption is predicted to be approximately 3.7 times as large as that of the Si-based modulator.

**Experiment**

The 6-inch Si/SiGe/Si-OI wafer was prepared by epitaxial growth on a commercially available (001) SOI wafer with a 2-μm-thick buried oxide (BOX). First, a 260-nm-thick SOI layer was thinned to be 100 nm by thermal oxidation. Then, a 30-nm-thick pseudomorphic $Si_{0.77}Ge_{0.23}$ layer and a 70-nm-thick Si layer were grown by chemical vapor deposition (CVD).

The *pin*-junction based optical modulator was fabricated by the conventional Si CMOS process. First, the straight rib waveguide with a Si/SiGe/Si core was formed by deep ultraviolet (DUV) lithography and dry etching. Figure 7(a) shows the cross-sectional TEM image of the Si/SiGe/Si rib waveguide with a 600-nm-wide mesa. The clear Si/SiGe/Si heterostructure is also observed in Fig. 7(b). Then, ion implantations of boron and phosphorus were carried out to make the $p^+$ and $n^+$ regions for forming a *pin*-junction, followed by activation annealing at 1000 $^{o}$C for 30 min in nitrogen atmosphere. Finally, the contact pads for the $p^+$ and $n^+$ regions were formed by thermal evaporation of aluminum. The Si-based device without a SiGe well was also fabricated by same process as a control device.

The amount of strain in the SiGe layer after the device fabrication was evaluated by Raman spectroscopy. It is well known biaxial strain partially relaxes when a biaxially strained layer is etched into a narrow mesa with sub-micro width such as an optical waveguide. Ge diffusion at the high-temperature activation annealing also causes strain reduction in the SiGe layer. Figure 7(c) shows the amount of strain in the SiGe measured before and after the dopant activation annealing at 1000 $^{o}$C for 30 min. The effect of the waveguide width on strain is also shown in Fig. 7(c) with changing the mesa width from 1900 nm to 600 nm. The dotted line in Fig. 7(c) shows the ideal strain curve of fully strained SiGe. The wide-mesa SiGe before annealing exhibits full strain, while the amount of strain decreases after annealing because the Ge fraction is reduced by Ge diffusion into the Si layers. The cross-section TEM of the Si/SiGe/Si heterostructure in the inset of Fig. 7(c) indicates that the thickness of the SiGe layer after annealing is increased to be approximately 50 nm. The dry etching of the 600-nm-wide waveguide also causes partial strain relaxation by 15% as shown in Fig. 7(c). Thus, the Ge fraction and compressive strain value of the SiGe layer after the device fabrication are approximately 14% and 0.48%, respectively, corresponding to 85%-strained $Si_{0.86}Ge_{0.14}$.

The optical attenuation was measured by injecting current for evaluating the enhancement of the free-carrier absorption in the strained SiGe layer. 1550 nm-wavelength continuous-wave (CW) TE-polarized light was coupled to the waveguides through a lensed fiber. Then, the output power was monitored by an InGaAs photodetector with changing the injection current.

**Result and Discussion**

Figure 8 shows the measured optical attenuation properties of the Si/SiGe/Si and Si in-line intensity modulators. The simulated optical attenuation properties, in which the changes in the Ge fraction and strain value after the device fabrication are taken into account, are also plotted as the solid lines. As shown in Fig. 8, the optical attenuation is increases by current injection through the *pin*-junction owing to the free-carrier absorption; the Si/SiGe/Si modulator exhibits higher attenuation at the same current density than that of the Si modulator. The experimental property of the Si/SiGe/Si devices shows a fairly good agreement with the simulation result. Therefore, it is clearly shown that strain-induced mass modulation enhances the free-carrier absorption and improves the device performance. The injected current densities for 20-dB attenuation are approximately 24 mA/mm and 55 mA/mm for the Si/SiGe/Si and the Si modulators, respectively. Thus, the modulation efficiency of the Si/SiGe/Si device is more than 2 times as large as that of the Si control device.

The wavelength dependences of the optical attenuations of the Si/SiGe/Si and Si modulators are also measured at the wavelength range from 1.34 to 1.64 μm. Since the free-carrier plasma dispersion effect is proportional to a square of wavelength according to the Drude model expressed by Equation (2), we plot the optical attenuations measured at injection currents of 20, 30, and 40 mA/mm as a function of a square of wavelength as shown in Fig. 9(a). The linear relationships between the optical attenuation and a square of wavelength are clearly observed, meaning that the optical attenuations arise from the free-carrier absorption. From the wavelength dependences of the optical attenuation characteristics, we calculated the effective refractive index change using Kramers-Kronig relationship. Since the wavelength range is limited from 1.34 to 1.64 μm corresponding to the photon energy range from 0.756 to 0.925 eV, the free-carrier absorption spectra are extrapolated into the far-infrared wavelength range. Figure 9(b) shows the calculated effective refractive index change at injection currents of 20, 30, and 40 mA/mm as a function of a square of wavelength as shown in Fig. 9(b), exhibiting the linear relationship of a square of wavelength.

The optical constants changes of the SiGe layer are deduced from the wavelength dependencies shown in Figs. 9 by taking into account optical confinement factor, strain relaxation and carrier concentration in the 50-nm-thick SiGe layer through the TCAD simulation. The optical constants changes of Si are also deduced by the same way. Figures 10 show the refractive index change and absorption coefficient change of 85%-strained $Si_{0.86}Ge_{0.14}$ and Si as a function of carrier concentration, in which the solid lines shows the theoretical values calculated by the Drude model. A fairly good agreement between the experiment results and the theory clearly indicates that the plasma dispersion effect and free-carrier absorption are enhanced by strain-induced mass modulation in strained SiGe. Thus, the enhancement factors of 1.3 for $\Delta n$ and 1.7 for $\Delta\alpha$ in 85%-strained $Si_{0.86}Ge_{0.14}$ have been successfully demonstrated.

## Conclusion

In conclusion, we have presented that the strain-induced enhancements of the plasma dispersion effect and free-carrier absorption in compressively strained SiGe is effective to boost the modulation efficiency of Si-based optical modulators. The optical attenuation of the SiGe-based in-line intensity optical modulator is more than 2 times larger than that of the Si modulator. To the best of our knowledge, it is the first demonstration of enhanced free-carrier absorption in SiGe through the strain-induced mass modulation for the telecommunication wavelength range from 1.3 to 1.6 μm. Since SiGe is already introduced into the CMOS production, the SiGe-based optical modulator presented here is highly CMOS compatible. Thus, the strained SiGe technology is easily applicable to most of the Si optical modulators based on the plasma dispersion effect and free-carrier absorption. We expect that the introduction of more strain by higher Ge fraction in SiGe enables further enhancement of the plasma dispersion effect and free-carrier absorption. Hence, the strain-induced enhancement of the plasma dispersion effect and free-carrier absorption is one of the most promising technology boosters for the Si-based optical modulators.

**References**

1 Soref, R. A. & Bennett, B. R. Electrooptical effects in silicon. *IEEE J. Quantum Electron.* **23**, 123–129(1987).

2 Xu, Q., Schmidt, B., Pradhan, S. & Lipson, M. Micrometre-scale Silicon electro-optic modulator. *Nature* **435**, (2005).

3 William, M. G., Michael, J. R., Lidija, S., & Yurii, A. V. Ultra-compact, low RF power, 10 Gb/s silicon Mach-Zehnder modulator. *Opt. Express* **15**, 17106–17113 (2007).

4 Sekiguchi, S., Kurahashi, T., Zhu, L., Kawaguchi, K. & Morito, K. Compact and low power operation optical switch using silicon-germanium/silicon hetero-structure waveguide. *Opt. Express* **20**, 17212–17219 (2012).

5 Gunn, C. CMOS Photonics for high-speed interconnects. *IEEE computer society*, **26**, 58–66, (2006).

6 Liao, L. *et al*. 40 Gbit/s silicon optical modulator for high-speed applications. *Electron. Lett*. **43**, (2007).

7 Gardes, F. Y., Thomson, D. J., Emerson, N. G. & Reed, G. T. 40 Gb/s silicon photonics modulator for TE and TM polarisations. *Opt. Express* **19**, 11804–11814 (2011).

8 Liu, A. *et al*. A high-speed silicon optical modulator based on a metal-oxide-semiconductor capacitor. *Nature* **427**, 615–618, (2004).

9 Liao, L. *et al*. High speed silicon Mach-Zehnder modulator. *Opt. Express* **13**, 3129–3135, (2005).

10 Liao, L. *et al*. Phase modulation efficiency and transmission loss of silicon optical phase shifters. *IEEE J. Sel. Top. Quantum Electron.* **41**, 250–257, (2005).

11 Dong、 P. *et al*. Wavelength-tunable silicon microring modulator. *Opt. Express* **18**, 10941–10946 (2010).

12 Nguyen, H. C., Sakai, Y., Shinkawa, M., Ishikura, N. & Baba, T. Photonic Crystal Silicon Optical Modulators: Carrier-Injection and Depletion at 10 Gb/s. *IEEE J. Quantum Electron*. **48**, 2, 210–220. (2012).

13 Baba, T. Slow light in photonic crystals. *Nature Photon*. **2**, 465–473 (2008).

14 Takenaka, M. & Takagi, S. Strain Engineering of Plasma Dispersion Effect for SiGe Optical Modulators. *IEEE J. Sel. Top. Quantum Electron.* **48**, 8–15 (2012).

15 Takagi, S., Sugiyama, N., Mizuno, T., Tezuka, T. & Kurobe, A. Device structure & electrical characteristics of strained-Si-on-insulator (strained-SOI) MOSFETs. *Materials Science and Engineering B* **89**, 426–434 (2002).

16 Rim, K. *et al*. Strained Si CMOS (SS CMOS) technology: opportunities and challenges. *Solid-State Electronics* **47**, 1133–1139 (2003).

17 Thompson, S. E., Sun, G., Choi, Y. S., C. & Nishida, T. Uniaxial-Process-Induced Strained-Si: Extending the CMOS Roadmap. *IEEE Trans. Electron Devices* **53**, 1010–1020 (2006).

18 Walton, A. K. Infrared Modulation and Energy Band Parameters in Multivalley Semiconductors through Uniaxial Stress Dependence of Free Carrier Contribution to Optical Constants. *phys. stat. sol. (b)* **43,** 379–386 (1971).

19 Walton, A. K. & Metcalfe, S. F. Free-carrier absorption at low temperatures in uniaxially stressed n-type Ge, Si and GaAs. *Solid State Phys.* **9**, 3605–3625 (1976)

20 Belyaev, A. E., Gorodnichii, O. P. & Pidlisny, E.V. Free carrier absorption in uniaxially stressed n-Si. *Solid State Communications* **42**, 441–445 (1982).

21 Douglas, J. P. Si/SiGe heterostructures: from material and physics to devices and circuits. *Semicond. Sci. Technol.* **19**, R75–R108 (2004).

22 People, R. & Bean, J. C. Calculation of critical layer thickness versus lattice mismatch for $Ge_xSi_{1-x}$/Si strained-layer heterostructures. *Appl. Phys. Lett.* **47**, 322–324 (1985).

23 J. M. Luttinger and W. Kohn, Motion of electrons and holes in perturbed periodic fields. *Phys. Rev.* **97**, 869–883 (1955).

24 Luttinger, J. M. Quantum theory of cyclotron resonance in semiconductors: General theory. *Phys. Rev.* **102**, 1030–1041 (1956).

25 Chao, C. Y. P. & Chuang, S. L. Spin-orbit-coupling effects on the valence-band structure of strained semiconductor quantum wells. *Phys. Rev. B* **46**, 4110–4122 (1992).

26 Martin, M. R. & Vogl, P. Electronic-band parameters in strained $Si_{1-x}Ge_x$ alloys on $Si_{1-y}Ge_y$ substrates. *Phys. Rev. B* **48**, 276–287 (1992).

27 Fischetti, M. V. & Laux, S. E. Band structure, deformation potentials, and carrier mobility in strained Si, Ge, and SiGe alloys. *J. Appl. Phys.* **80**, 2234–2252 (1996).

28 Arora, N. D., Hauser, J. R. & Roulston, D. J. Electron and hole mobilities in silicon as a function of concentration and temperature. *IEEE Trans. Electron Devices* **29**, 292–295 (1982).

29 Schaffler, F. *Properties of Advanced Semiconductor Materials GaN, AlN, InN, BN, SiC, SiGe*. New York: Wileygk (2001).

30 Gui-Rong, Z. *et al*. Effect of carrier lifetime on forward-biased silicon Mach-Zehnder modulators. *Opt. Express* **16**, 5218–5226 (2008).

**Acknowledgment**

This work was partly supported by the Strategic Information and Communications R&D Promotion Programme of the Ministry of Internal Affairs and Communications.

**Author contributions**

Y. K. proposed the device design and process, designed the experiment, fabricated the samples, collected the data, and performed analysis of data; T. O., and M. H. fabricated the epitaxial Si/SiGe on Si-on-isulator semiconductor substrates; M. T. and S. T. managed the research and supervised the experiment; All authors discussed the results and commented on the manuscript.

## Figures

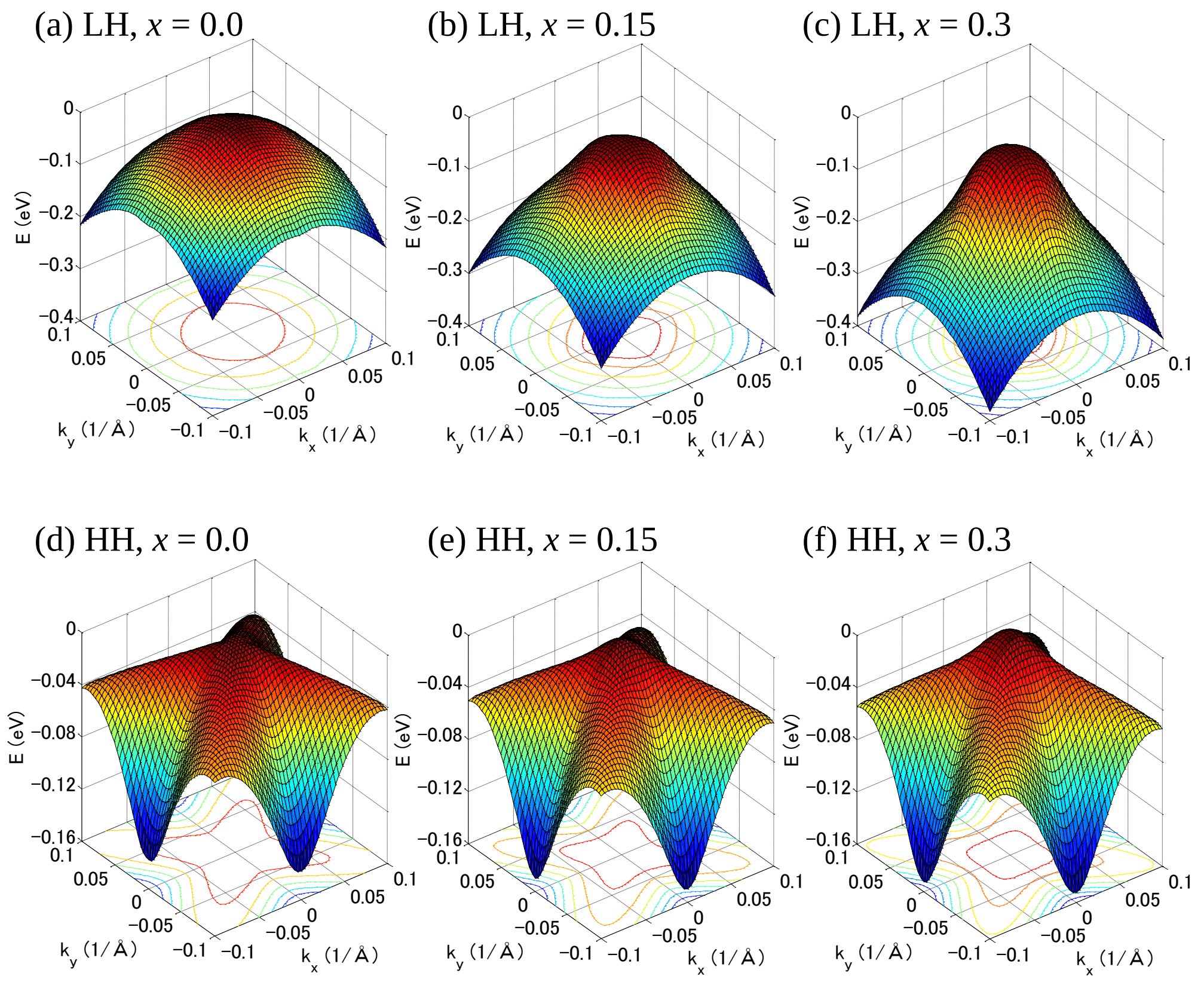

**FIG. 1. Valence band structures and constant-energy contour of $Si_{1-x}Ge_x$ grown on Si (001) as a function of wavevectors in the $k_x$ [100]-*ky* [010] plane calculated by six-band *kp* method.** Top three figures (a), (b) and (c) are LH band and bottom three figures (d), (e) and (f) are HH band at x = 0 (Si), x = 0.15 and 0.3 (SiGe). The energy surfaces of the LH and HH bands become sharp at the minimum energy Γ point with an increase in Ge fraction, indicating the biaxial compressive strain reduces the effective hole masses.

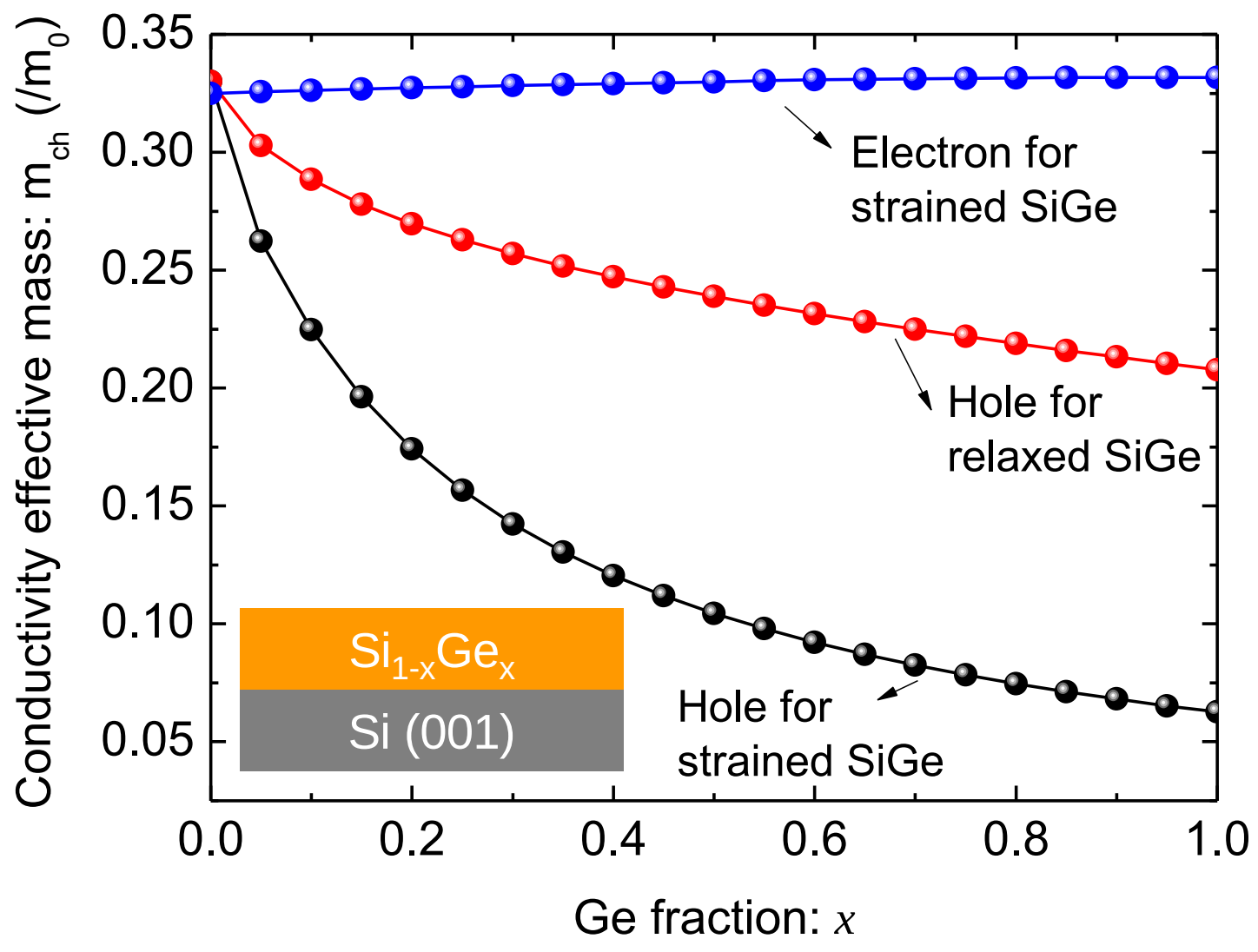

**FIG. 2. Calculated in-plane conductivity effective hole masses of compressively strained $Si_{1-x}Ge_x$ grown on Si (black) and relaxed $Si_{1-x}Ge_x$ (red), and electron masses for strained SiGe[26] (blue).** Comparing between the hole masses of relaxed and strained SiGe, the compressive strain has a large effect on reducing the conductivity hole mass.

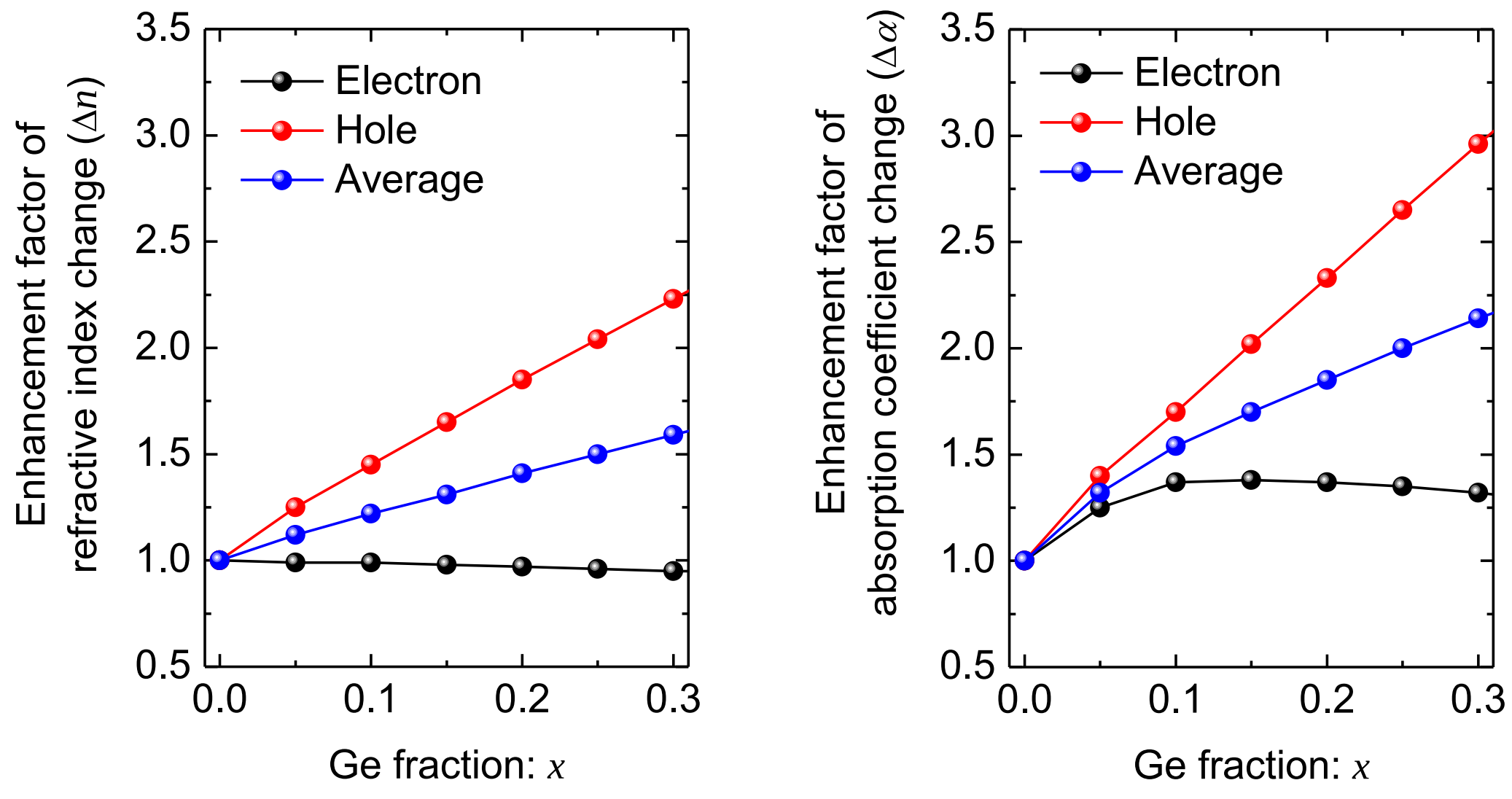

(a) (b)

**FIG. 3. Enhancement factors of optical constants changes.** (a) refractive index and (b) absorption coefficient as function of Ge fraction. The optical constants changes increase with an increase in Ge fraction and are expected to be enhanced as large as 1.6 and 2.1 for $\Delta n$ and $\Delta\alpha$ respectively in compressively strained $Si_{0.7}Ge_{0.3}$.

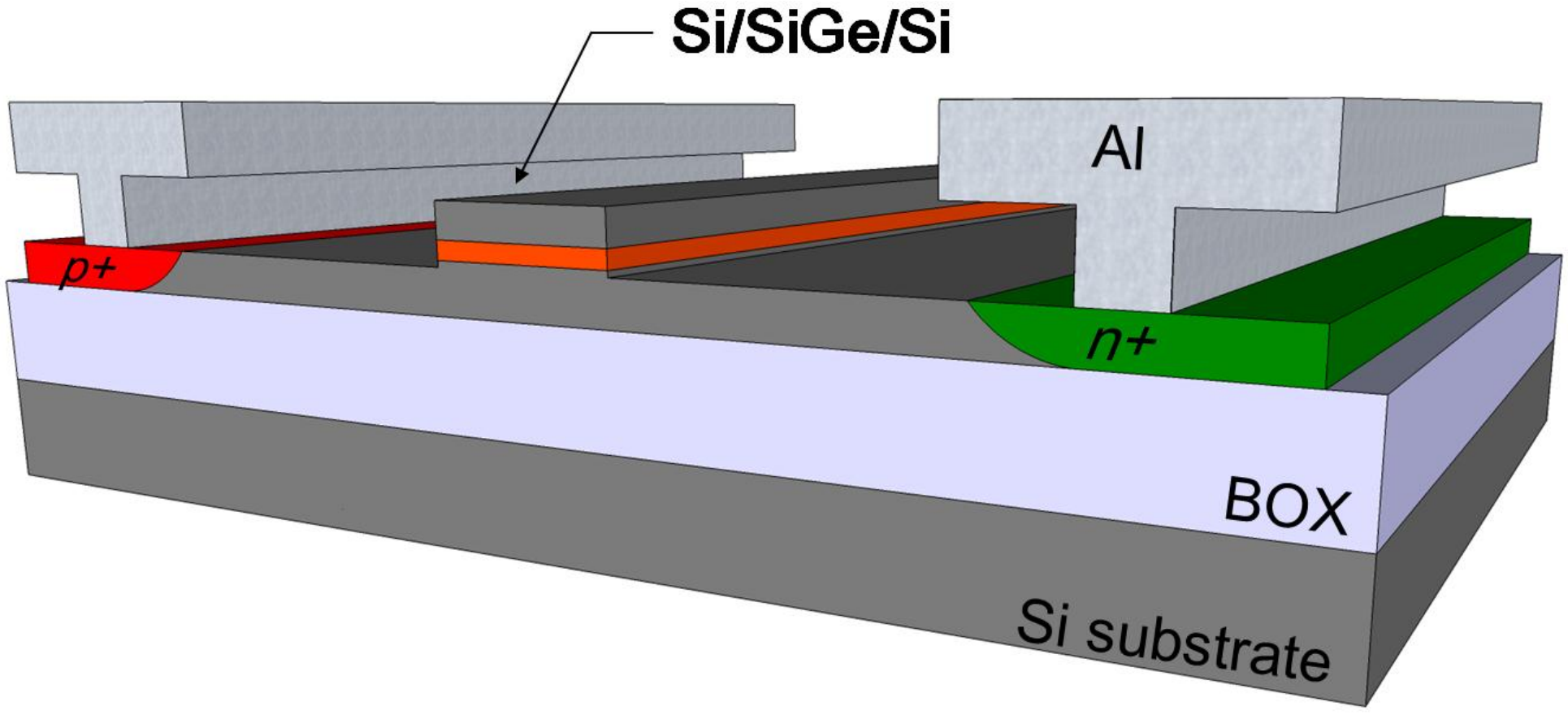

(a)

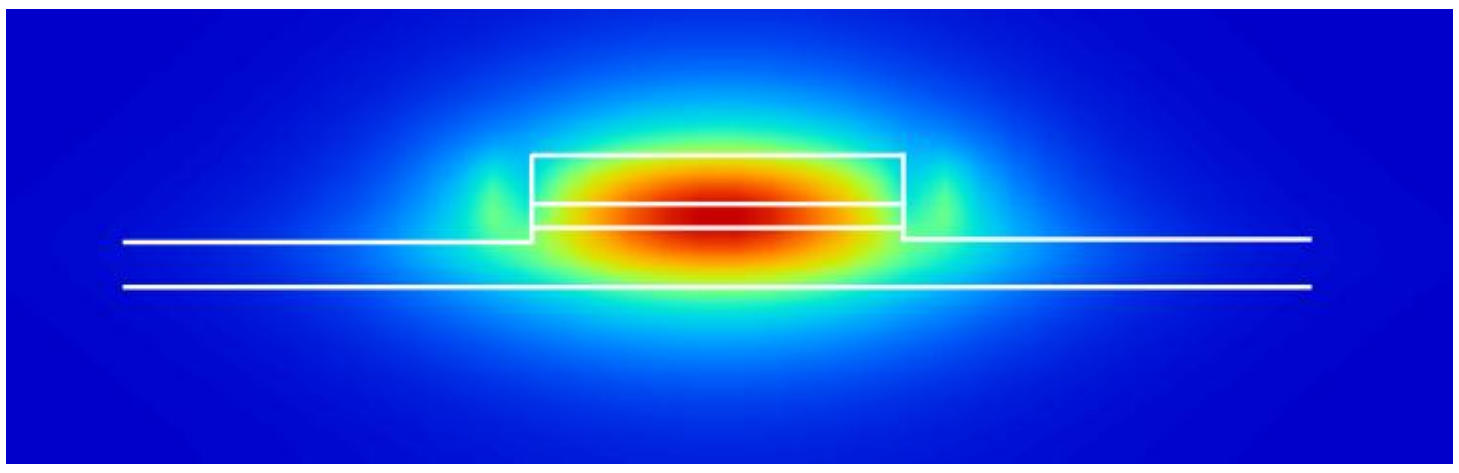

(b)

**FIG. 4. Device structure of optical modulator.** (a) Device structure of the *pin* injection-type optical modulator with Si/SiGe/Si waveguide core. (b) Electric field distribution of the fundamental TE mode of the modulator.

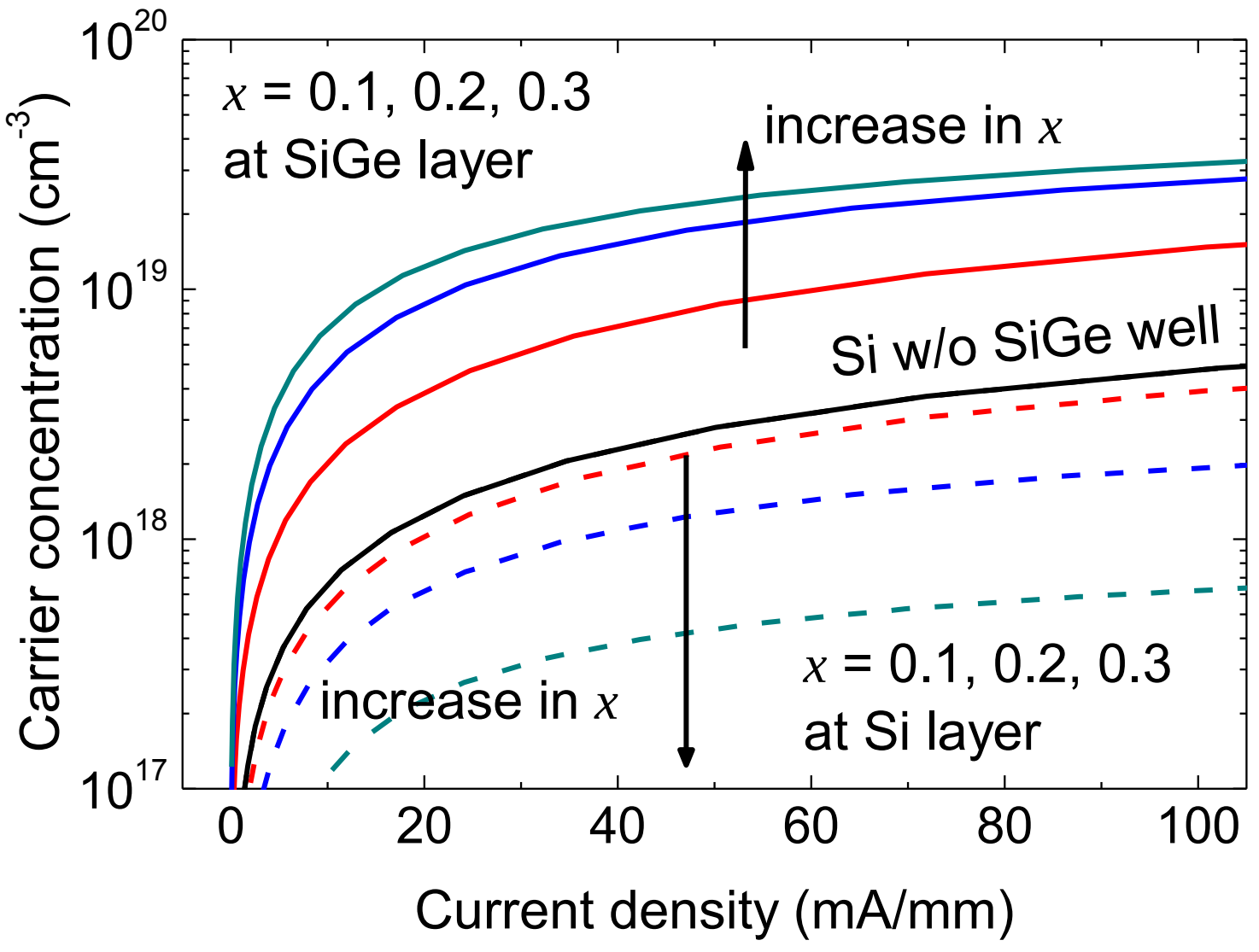

(a)

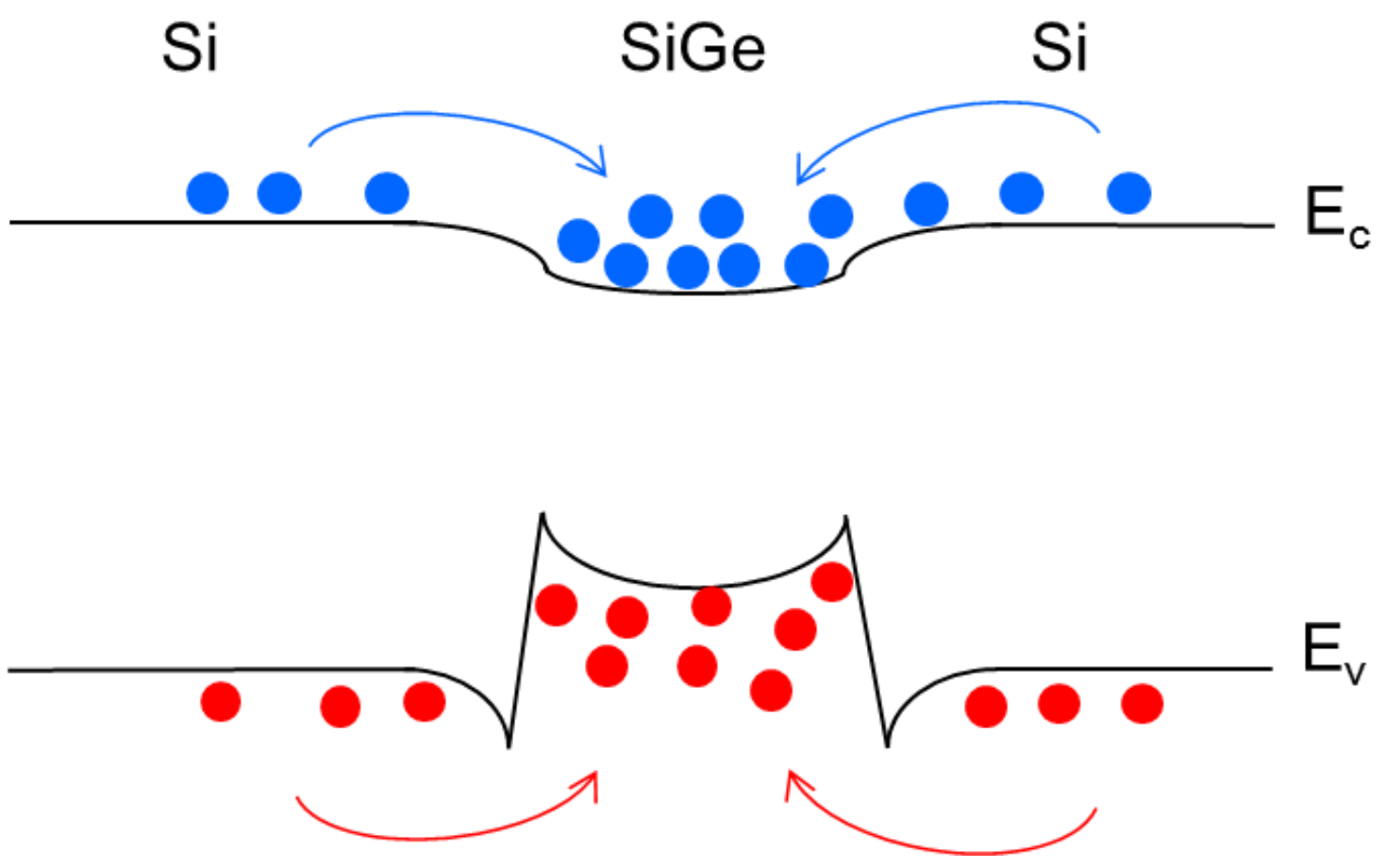

(b)

**FIG. 5. Carrier confinement in SiGe layer.** (a) Carrier concentrations at Si and SiGe layer as a function of injected current density with Ge fractions, 0.0 to 0.3. (b) Bandstructure of Si/SiGe/Si heterostructure illustrating carrier confinement.

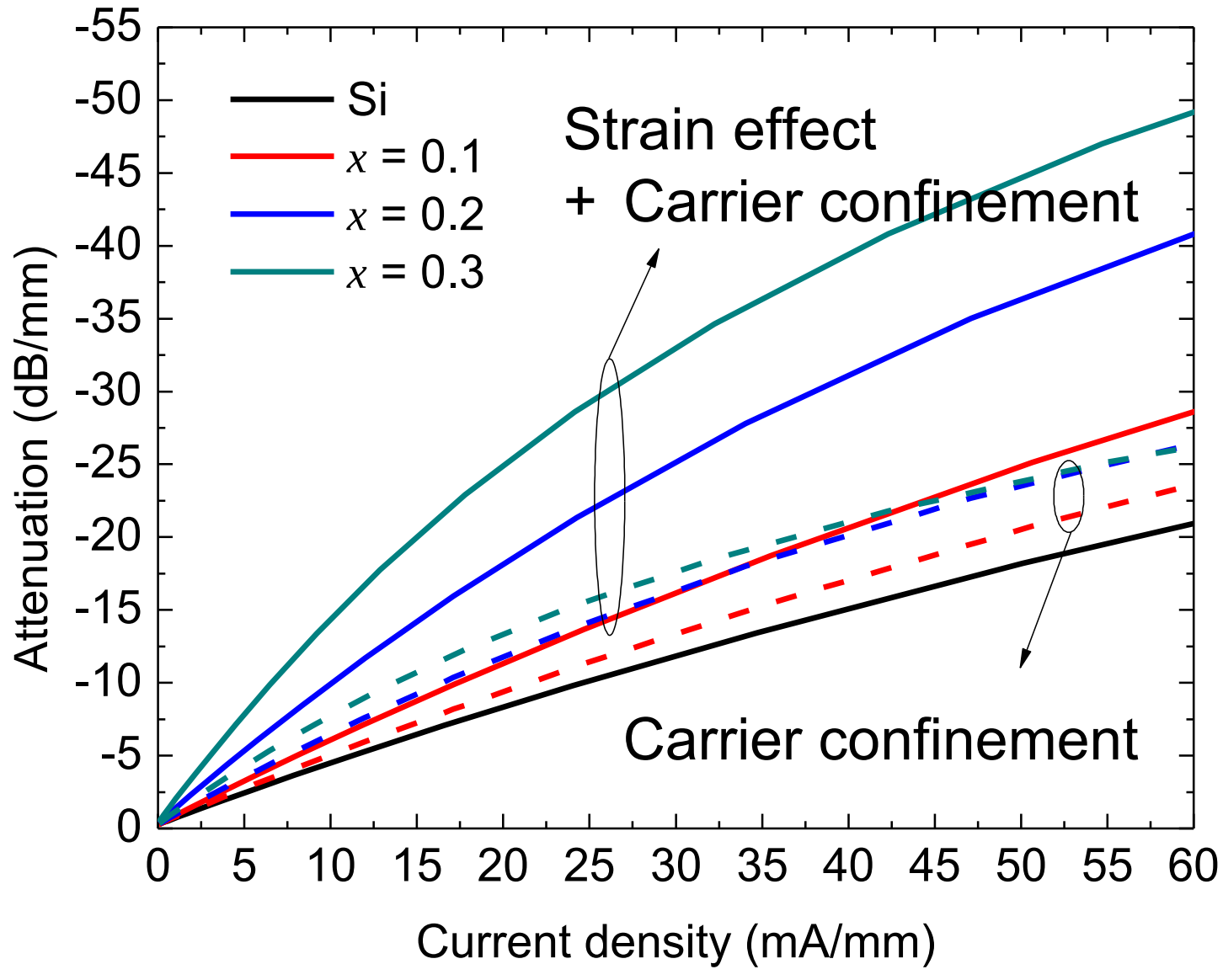

-55
-50
-45
-40
-35
-30
-25
-20
-15
-10
-5
0
Attenuation (dB/mm)
Si
x = 0.1
x = 0.2
x = 0.3
Strain effect
+ Carrier confinement
Carrier confinement
0 5 10 15 20 25 30 35 40 45 50 55 60
Current density (mA/mm)

(a)

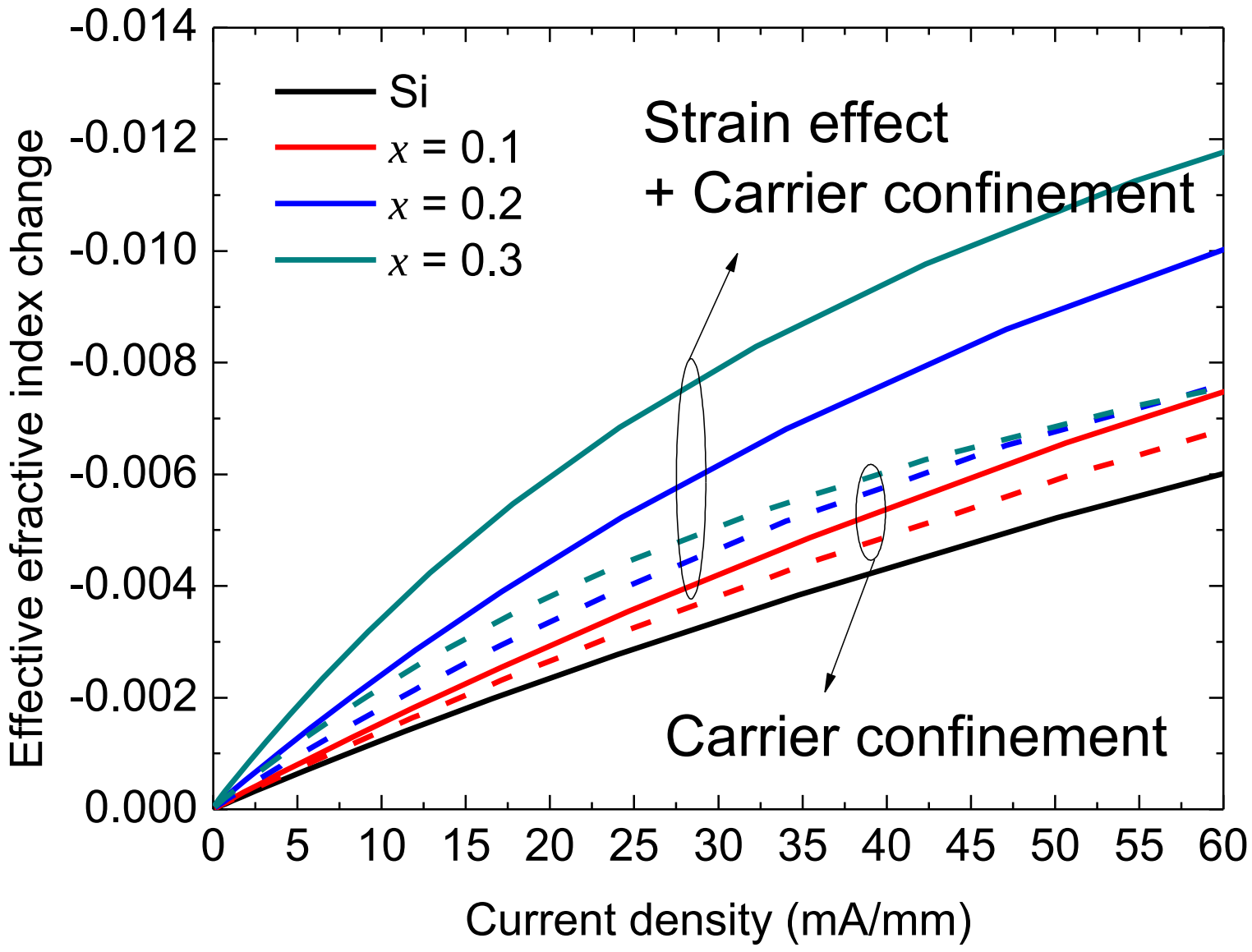

-0.014
-0.012
-0.010
-0.008
-0.006
-0.004
-0.002
0.000
Effective efractive index change
Si
x = 0.1
x = 0.2
x = 0.3
Strain effect
+ Carrier confinement
Carrier confinement
0 5 10 15 20 25 30 35 40 45 50 55 60
Current density (mA/mm)

(b)

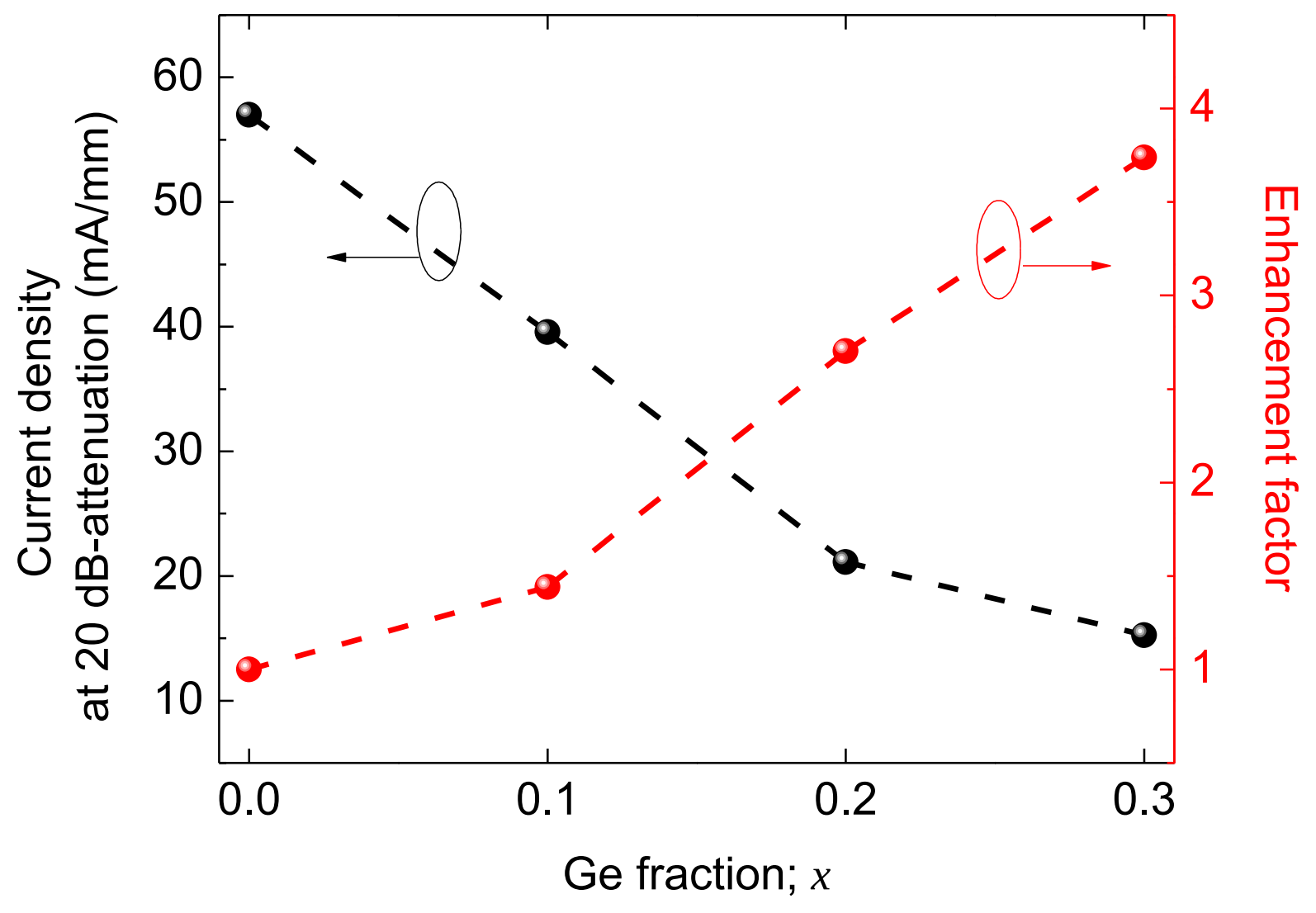

(c)

**FIG. 6. Simulation results of the *pin* injection-type optical modulator with Si/SiGe/Si waveguide core.** (a) Attenuation characteristics as a function of injected current density to in-line modulator with Ge fractions, 0.0 to 0.3. (b) Refractive index changes as a function of injected current density to in-line modulator with Ge fractions, 0.0 to 0.3. (c) Current density required for 20 dB-attenuation (left-axis) and its enhancement factor (right-axis) as a function of Ge fractions; the efficiency of the $Si_{0.7}Ge_{0.3}$-based in-line intensity modulator is predicted to be approximately 3.7 times as large as that of the Si-based modulator.

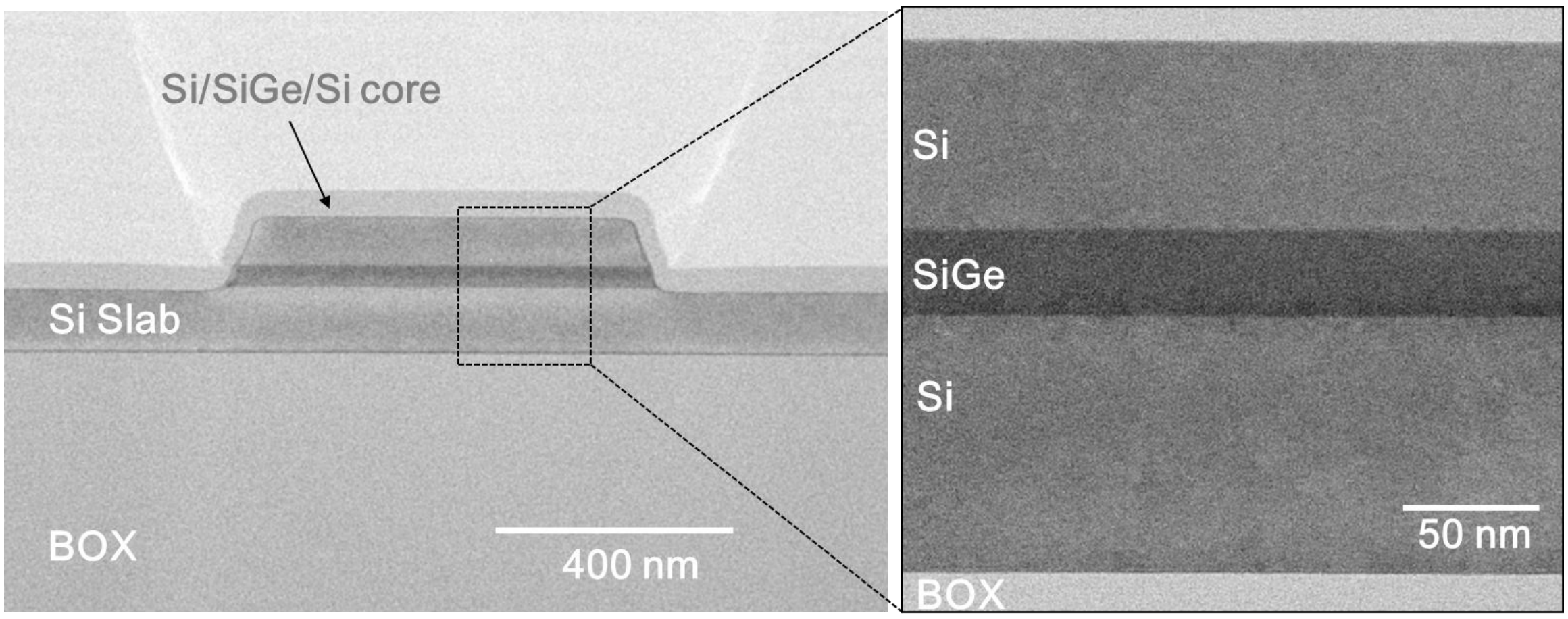

(a) (b)

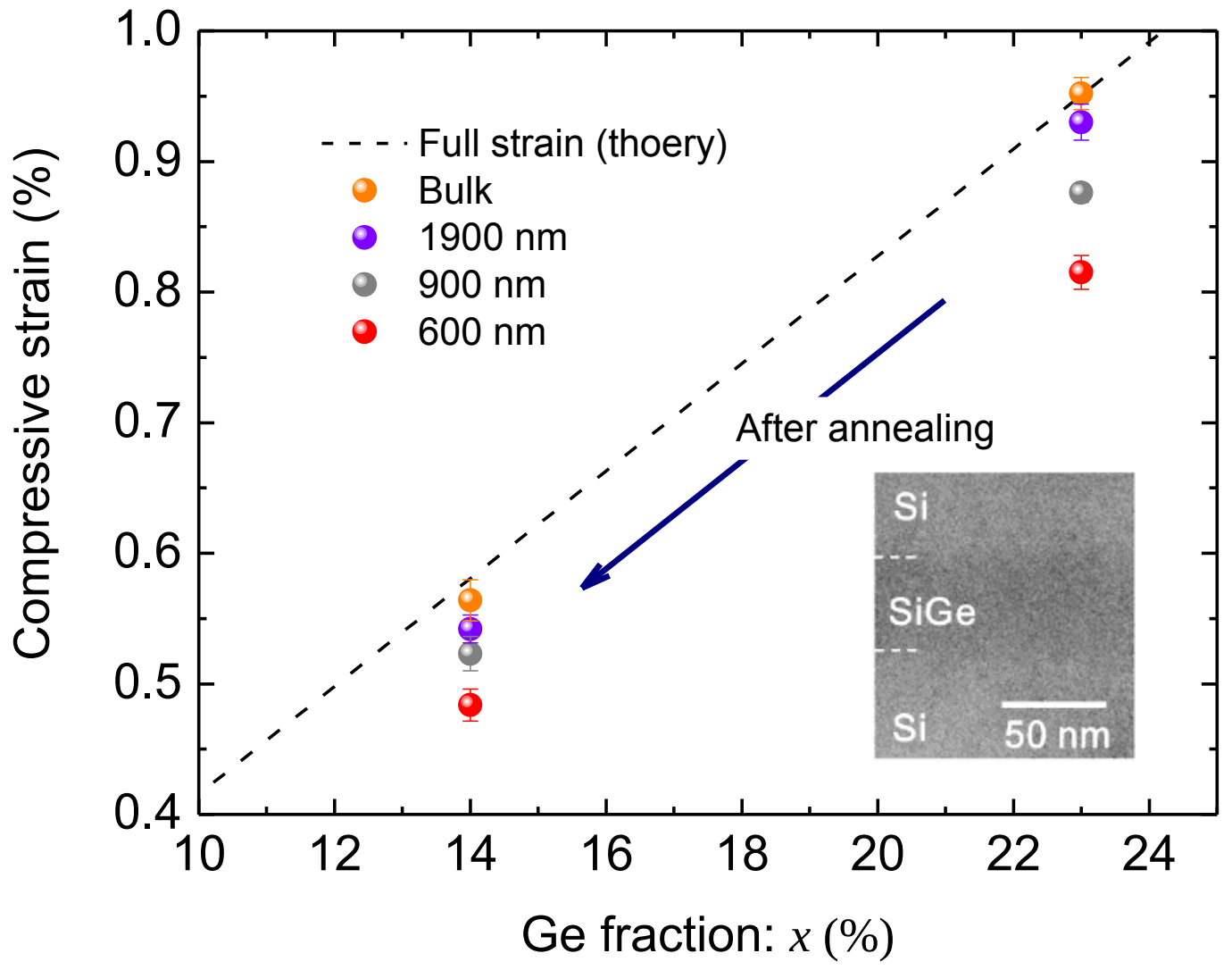

(c)

**FIG. 7 TEM images of the fabricated device and evaluation of strain and Ge fraction.** (a) Cross-sectional TEM image of Si/SiGe/Si-core waveguide and (b) Si/SiGe/Si heterostructure. (c) Compressive strain as a function of Ge fraction with waveguide widths and cross-sectional TEM image of Si/SiGe/Si heterostructure after annealing process (inset).

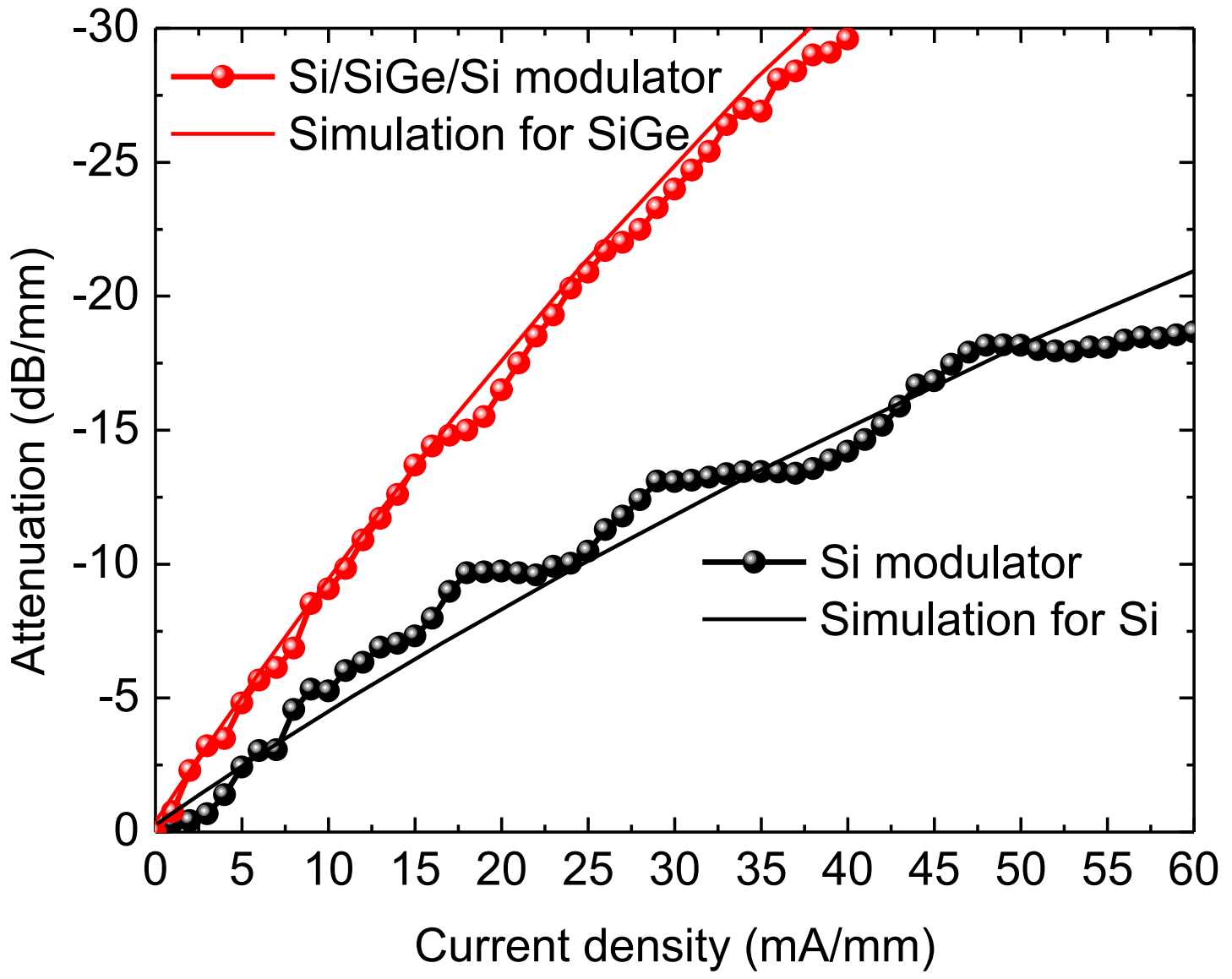

**FIG. 8. Attenuation characteristics of Si/SiGe/Si and Si modulator**. Experimental results are shown by dotted line and simulated results are shown by line. The modulation efficiency of the Si/SiGe/Si device is more than 2 times as large as that of the Si control device.

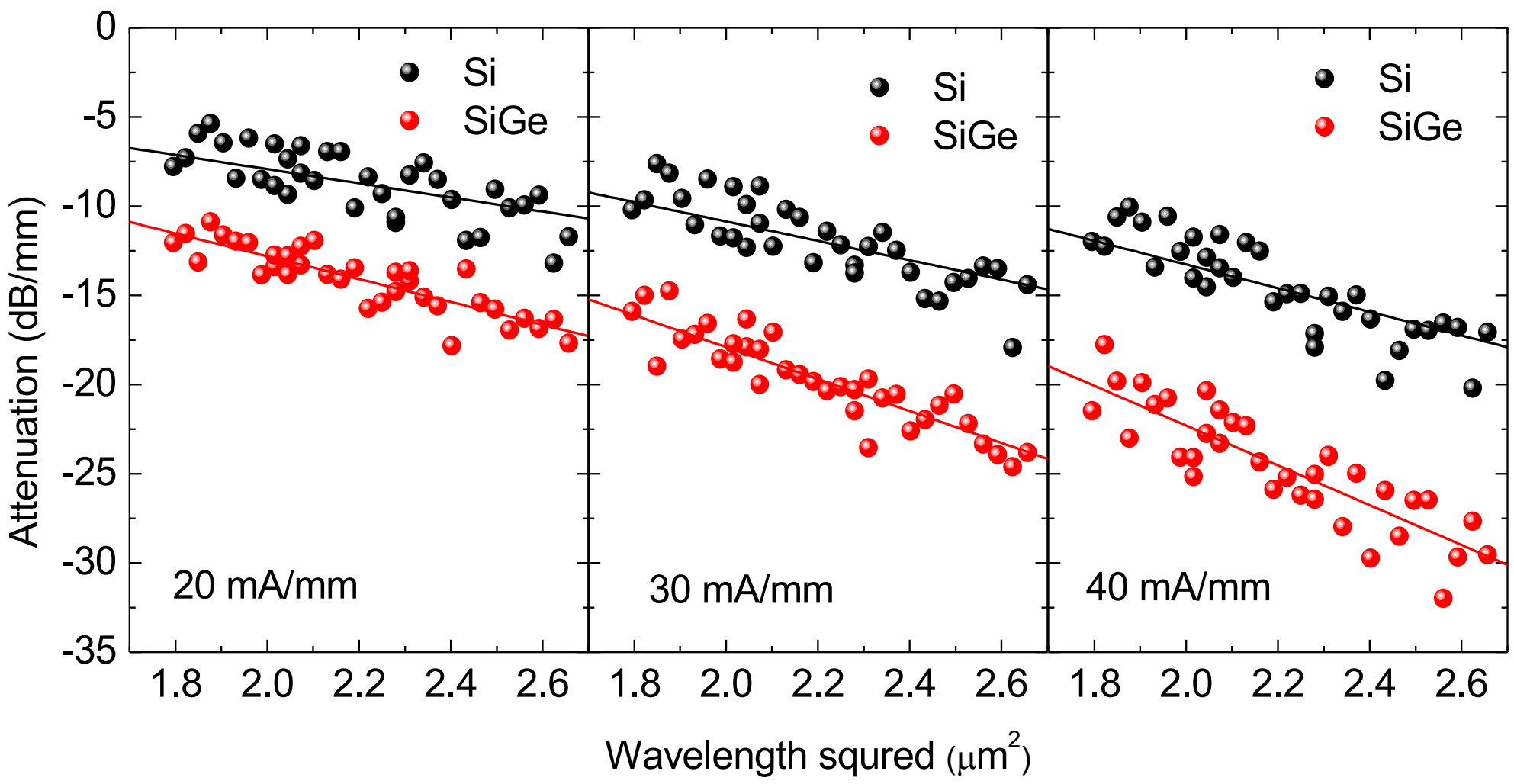

(a)

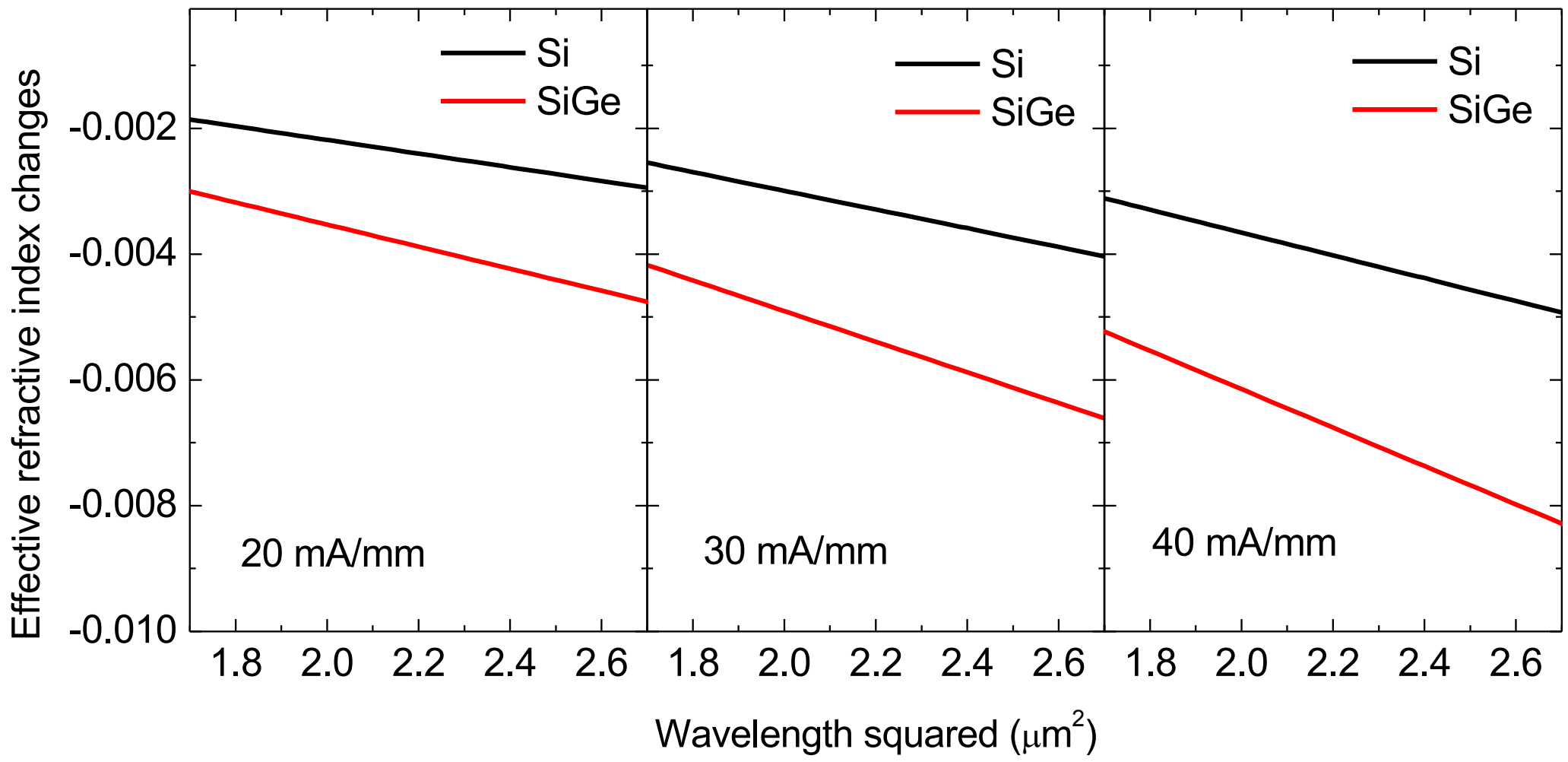

(b)

**FIG. 9. Wavelenth dependence of modulation characteristics.** (a) Attenuation measured by Si/SiGe/Si and Si modulator with current densities of 20, 30 and 40 mA/mm. (b) Effective refractive index change calculated by Kramers-Kronig relationship from measured wavelength dependence of attenuation.

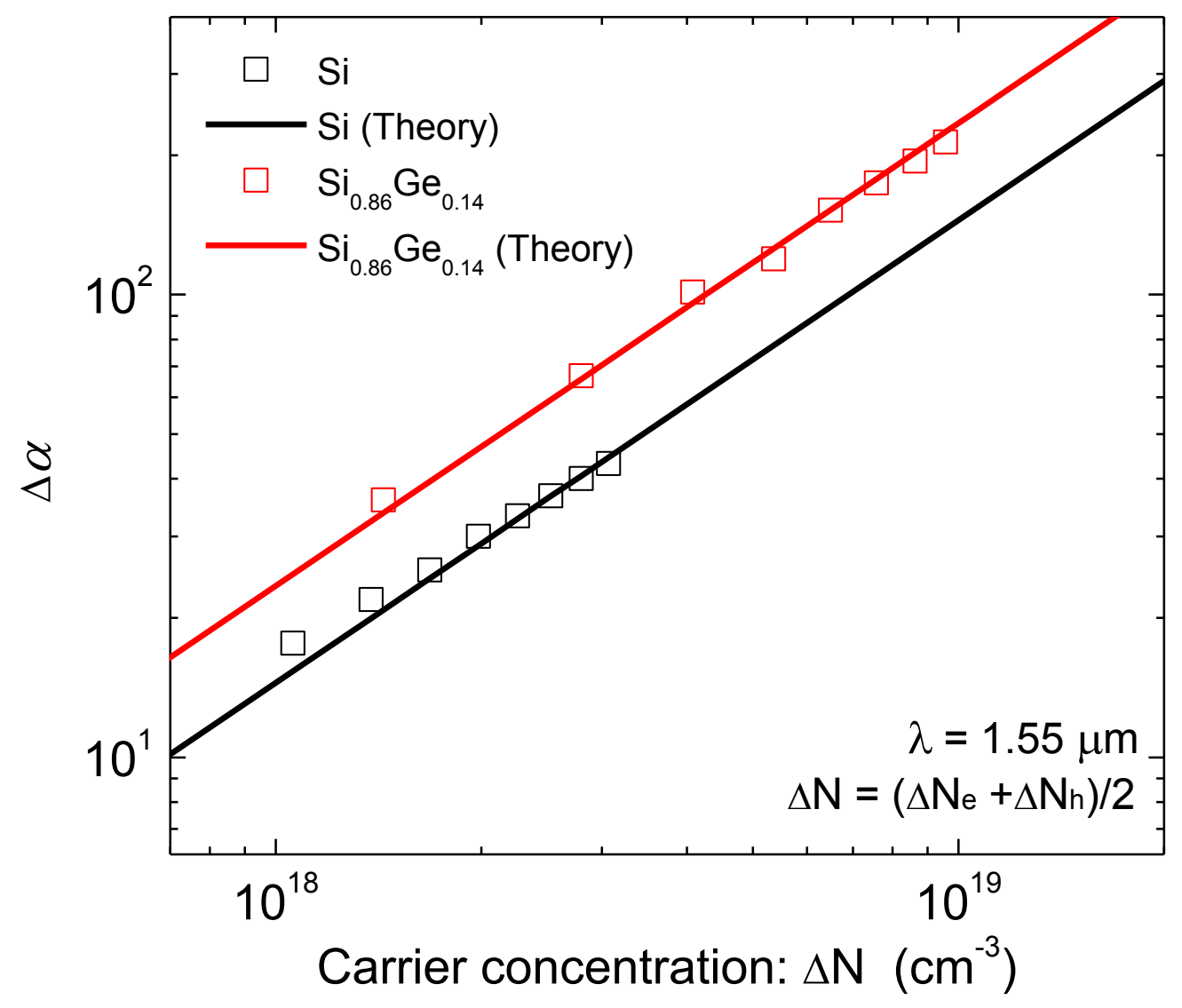

(a)

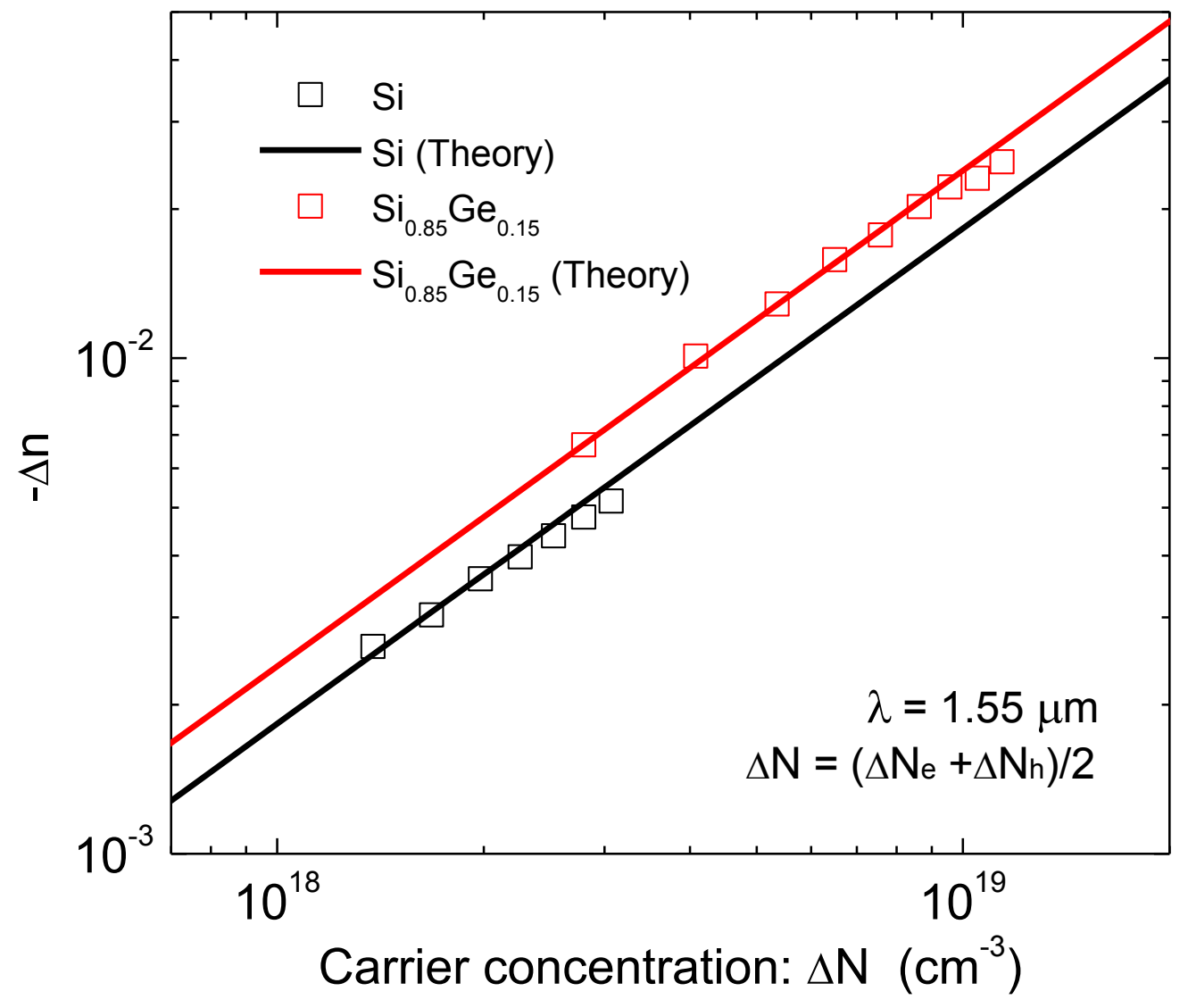

(b)

**FIG. 10. Optical constants change as a function of carrier concentration.** (a) absorption coefficient change and (b) refractive index change. The experiment results and the theory have a good agreement indicating the plasma dispersion effect and free-carrier absorption are enhanced by strain-induced mass modulation in strained SiGe.